\documentclass[aps,prd,twocolumn,groupedaddress,showpacs,superscriptaddress,floatfix]{revtex4-1}

\usepackage{amsmath}
\usepackage{graphicx}
\usepackage{amssymb}
\usepackage{bm}
\usepackage{array}
\usepackage[colorlinks,linkcolor=blue,urlcolor=blue,anchorcolor=blue,citecolor=blue]{hyperref}
\usepackage{slashed}
\usepackage{braket}

\begin{document}

\title{Angular momentum decomposition from a QED example}

\newcommand*{\PKU}{School of Physics and State Key Laboratory of Nuclear Physics and
Technology, Peking University, Beijing 100871,
China}\affiliation{\PKU}
\newcommand*{\INFN}{INFN, Sezione di Pavia, via Bassi 6, 27100 Pavia, Italy}\affiliation{\INFN}
\newcommand*{\CICQM}{Collaborative Innovation
Center of Quantum Matter, Beijing, China}\affiliation{\CICQM}
\newcommand*{\CHEP}{Center for High Energy
Physics, Peking University, Beijing 100871,
China}\affiliation{\CHEP}

\author{Tianbo Liu}\email{liutb@pku.edu.cn}\affiliation{\PKU}\affiliation{\INFN}
\author{Bo-Qiang Ma}\email{mabq@pku.edu.cn}\affiliation{\PKU}\affiliation{\CICQM}\affiliation{\CHEP}


\begin{abstract}
We investigate the angular momentum decomposition with a quantum electrodynamics example to clarify the proton spin decomposition debates. We adopt the light-front formalism where the parton model is well defined. We prove that the sum of fermion and boson angular momenta is equal to half the sum of the two gravitational form factors $A(0)$ and $B(0)$, as is well known. However, the suggestion to make a separation of the above relation into the fermion and boson pieces, as a way to measure the orbital angular momentum of fermions or bosons, respectively, is not justified from our explicit calculation.
\end{abstract}


\maketitle

Angular momentum decomposition of a relativistic composite system is a fundamental problem in physics and is one of the most active frontiers in recent years. Although the total angular momentum of an isolated system is well defined, its decomposition into the spin and orbital angular momentum of each component in interaction theories is nontrivial and of great interest. In quantum chromodynamics (QCD) where we have no free quarks or gluons, it is challenging to relate each term in the decomposition, especially the orbital angular momentum (OAM), to physical observables. This is of physical significance in understanding the nucleon structure.

The famous ``proton spin crisis,'' i.e., the observation that the quark spin only contributes a small fraction~\cite{Ashman:1987hv} (about 30\% from recent analyses~\cite{Ageev:2005gh}) to the proton spin, puzzled the whole physics society. This result severely deviates from the naive quark model. A straightforward understanding is to attribute the remaining proton spin to OAM and/or gluon helicity. Because of the Wigner rotation~\cite{Wigner:1939cj} that relates spin states between instant form and light-front form (or
those between rest frame and infinite momentum frame)~\cite{Ma:1991xq}, one can obtain a nonvanishing OAM contribution from the extension of a nonrelativistic $s$-wave quark model to a relativistic light-front treatment~\cite{Ma:1998ar}. Therefore, the measurement of the OAM is important, though the gluon may also contribute a large portion to the proton spin~\cite{deFlorian:2014yva}. However, the definition of OAM in a gauge field theory is still under debate.

A most intuitive decomposition, given by Jaffe and Manohar~\cite{Jaffe:1989jz}, breaks the gauge invariance and therefore seems unmeasurable. Then a manifest gauge invariant decomposition was proposed~\cite{Ji:1996ek}, and in this decomposition the total angular momentum of each parton flavor is related to the sum of two gravitational form factors, $A(0)$ and $B(0)$, which can be measured through deeply virtual Compton scattering (DVCS) processes. These kinds of relations can shed light on the measurement of OAM if they are generally valid, but we will find in this report that this relation is unjustified.

By splitting the gauge potential into pure gauge and physical terms, Chen {\it et al}. suggested a decomposition~\cite{Chen:2008ag} in which the operators of each term are gauge invariant and satisfy the angular momentum commutation relations. Then many decomposition versions are proposed with this approach~\cite{Wakamatsu:2010cb}. In these decompositions, it seems that a special gauge in which the pure gauge term vanishes is still implied. In fact, it comes from the so-called Stueckelberg symmetry, which copies the group of gauge symmetry but acts on the fields in a different manner, in separating the pure gauge term. Hence the approach of Chen {\it et al}. can be viewed as a gauge invariant extension (GIE) based on a Stueckelberg symmetry fixing~\cite{Lorce:2012rr}. This fixing procedure is essentially a choice of the separation for the pure gauge and physical terms, and thus may result in different decomposition versions which actually correspond to different physical objects~\cite{Wakamatsu:2014toa}. Because of the longitudinal boost invariance, the light-front gauge motivated choice is favored by the parton language. Nowadays, all of the decompositions are usually divided into two classes~\cite{Leader:2013jra}, the canonical class, e.g., Jaffe and Manohar's (JM's), and the kinetic or mechanical class, e.g., Ji's. Both of them share the same term for fermion spin, and the main difference between them is the definition of fermion OAM. From the GIE procedure, the OAMs defined in both classes are, in principle, measurable without gauge invariance breaking, but the connection of the OAM to physical observables is still open to challenge.

Recently, some quark model calculations show that canonical and kinetic OAMs are different even in no gauge field models~\cite{Liu:2014npa,Lorce:2011kd}, in which cases one believes that the two definitions should coincide with each other and give the same results. In this report, we perform explicit calculations in quantum electrodynamics (QED) in order to avoid model assumptions. Since perturbative calculations with QED are quite reliable and have been precisely tested, it is an ideal theoretical laboratory to test the consistency of the formulas. Besides, the decompositions derived in QCD also work in QED and have the same structure for each term. In fact, the QED structure of an electron has been discussed in the literature~\cite{Brodsky:1980zm,Brodsky:2000ii,Burkardt:2008ua} to clarify issues concerning the spin structure of a composite system, and our work can be considered a natural extension in that tradition. Since the parton model is established in an infinite momentum frame and the constituents are unambiguously defined in light-front quantization~\cite{Brodsky:1997de}, we perform our calculations in light-front form.

We start from the QED Lagrangian
\begin{equation}
\begin{split}
\mathcal{L}=&\frac{i}{2}\left[\bar{\psi}\gamma^\mu\partial_\mu\psi-(\partial_\mu\bar{\psi})\gamma^\mu\psi\right]-m\bar{\psi}\psi\\
&-e\bar{\psi}\gamma^\mu A_\mu\psi-\frac{1}{4}F_{\mu\nu}F^{\mu\nu}.
\end{split}
\end{equation}
Noether's theorem leads to the expression for the canonical energy-momentum tensor~\cite{Noether:1918zz}. It differs from the Belinfante improved energy-momentum tensor by a divergence term of the so-called superpotential~\cite{Belinfante1939}. One can formally divide the energy-momentum tensor into three parts,
\begin{equation}
T^{\mu\nu}=T^{\mu\nu}_f+T^{\mu\nu}_b+T^{\mu\nu}_I,
\end{equation}
where the subscripts $f$, $b$, and $I$ represent the fermion, boson, and interaction parts. Their expressions are
\begin{align}
T^{\mu\nu}_f&=\frac{i}{2}\left[\bar{\psi}\gamma^\mu\partial^\nu\psi-(\partial^\nu\bar{\psi})\gamma^\mu\psi\right]\label{emf}\\
&\quad-\frac{i}{2}g^{\mu\nu}\left[\bar{\psi}\gamma^\rho\partial_\rho\psi-(\partial_\rho\bar{\psi})\gamma^\rho\psi\right]+g^{\mu\nu}m\bar{\psi}\psi,\nonumber\\
T^{\mu\nu}_b&=-F^{\mu\rho}\partial^\nu A_\rho+\frac{1}{4}g^{\mu\nu}F^{\rho\sigma}F_{\rho\sigma},\label{emb}\\
T^{\mu\nu}_I&=g^{\mu\nu}e\bar{\psi}\gamma^\rho A_\rho\psi.\label{emi}
\end{align}
The interaction term Eq.~(\ref{emi}) is usually absorbed into the fermion part by using the equation of motion:
\begin{equation}
\left(i\gamma^\mu\partial_\mu-m-e\gamma^\mu A_\mu\right)\psi=0.
\end{equation}
Actually, one may also attribute it to the boson part through the other equation of motion, the Maxwell equation:
\begin{equation}
\partial_\mu F^{\mu\nu}-e\bar{\psi}\gamma^\nu\psi=0.
\end{equation}
Therefore, the decomposition of the energy-momentum tensor into a fermion part and a boson part is not unique or, in other words, artificial.

However, similar to the Dirac and Pauli electromagnetic form factors which correspond to the helicity-conserving and helicity-flip matrix elements of the plus component $J^+$ of the vector current in the light-front formalism~\cite{Brodsky:1980zm}, the gravitational form factors $A(Q^2)$ and $B(Q^2)$ are identified from the matrix elements of tensor component $T^{++}$ as~\cite{Brodsky:2000ii}
\begin{align}
\left\langle P+q,\uparrow\left|T^{++}(0)\right|P,\uparrow\right\rangle&=2(P^+)^2A(Q^2),\label{gffa}\\
\left\langle P+q,\uparrow\left|T^{++}(0)\right|P,\downarrow\right\rangle&=2(P^+)^2\frac{-(q^1-iq^2)}{2m}B(Q^2),\label{gffb}
\end{align}
where $q$ is a spacelike four-momentum with invariant mass square $q^2=-Q^2$. One may easily find that the interaction term $T^{++}_I$ vanishes, and thus we are able to uniquely separate the fermion and boson contributions to the form factors via $T^{++}_f$ and $T^{++}_b$:
\begin{align}
T^{++}_f&=\frac{i}{2}\left[\bar{\psi}\gamma^+\partial^+\psi-(\partial^+\bar{\psi})\gamma^+\psi\right],\label{emfplus}\\
T^{++}_b&=-F^{+\rho}\partial^+ A_\rho=\partial^+A^j\partial^+A^j,\label{embplus}
\end{align}
where the light-front gauge $A^+=0$ is adopted.

Quantized at fixed light-front time $\tau=(t+z)/\sqrt{2}$, the fermion and boson field operators are expanded as~\cite{Brodsky:1997de}
\begin{align}
\psi(x)=&\sum_\lambda\int\frac{d\ell^+d^2\vec{\ell}_\perp}{\sqrt{2\ell^+}(2\pi)^3}[b_\lambda(\ell)u_\lambda(\ell)e^{-i\ell\cdot x}\nonumber\\
&+d_\lambda^\dagger(\ell)v_\lambda(\ell)e^{i\ell\cdot x}],\\
A^j(x)=&\sum_\lambda\int\frac{d\ell^+d^2\vec{\ell}_\perp}{\sqrt{2\ell^+}(2\pi)^3}[a_\lambda(\ell)\epsilon^i_\lambda(\ell)e^{-i\ell\cdot x}\nonumber\\
&+a_\lambda^\dagger(\ell)\epsilon^{i*}_\lambda(\ell)e^{i\ell\cdot x}],
\end{align}
with the commutation and anticommutation relations of the creation and annihilation operators:
\begin{equation}
\begin{split}
&[a_\lambda(q),a_{\lambda'}(q')]=\{b_\lambda(q),b_{\lambda'}(q')\}\\
=&(2\pi)^3\delta(q^+-q'^+)\delta^{(2)}(\vec{q}_\perp-\vec{q}'_\perp)\delta_{\lambda\lambda'},
\end{split}
\end{equation}
where $\lambda$ is the light-front helicity. The one particle state is defined as $\sqrt{2q^+}\,a^\dagger(q)|0\rangle$. We adopt the Lepage-Brodsky convention for Dirac spinors $u$, $v$ and polarization vectors $\epsilon$~\cite{Lepage:1980fj}.

In the language of light-front quantization in QED, the physical electron state is expanded on a complete basis of Fock states composed of fermions and gauge bosons~\cite{Brodsky:1997de}. We cut off the expansion to two-particle Fock state. This corresponds to the one-loop level in Feynman diagram language. Then the physical electron state with momentum $P$ and helicity $S_z$ is expressed as
\begin{equation}
\begin{split}
&\left|P^+,\vec{P}_\perp,S_z\right\rangle\\
&=\sqrt{Z}\sqrt{2P^+}b_{S_z}^\dagger(P^+,\vec{P}_\perp)|0\rangle\\
&\quad +\sum_{\sigma,\lambda}\int\frac{dxd^2\vec{k}_\perp}{2(2\pi)^3}2P^+\psi^{S_z}_{\sigma,\lambda}(x,\vec{k}_\perp)b_\sigma^\dagger(xP^+,x\vec{P}_\perp+\vec{k}_\perp)\\
&\quad\quad a_\lambda^\dagger((1-x)P^+,(1-x)\vec{P}_\perp-\vec{k}_\perp)|0\rangle,\label{fockexp}
\end{split}
\end{equation}
where $Z$ is the renormalization constant or one-particle Fock state light-front wave function, and $\psi^{S_z}_{\sigma,\lambda}(x,\vec{k}_\perp)$ is the two-particle Fock state light-front wave function. The $x$ and $\vec{k}_\perp$ are the light-front momentum fraction and the intrinsic transverse momentum carried by the fermion. Together with the field operator expansions, after some algebra, the matrix elements of local operators can be represented in terms of light-front wave functions.

In QED, the wave functions can be systematically perturbatively evaluated~\cite{Lepage:1980fj}. An electron with $S_z=+1/2$ has four possible helicity combinations for a two-particle Fock state~\cite{Brodsky:2000ii}. Their light-front wave functions are
\begin{align}
\psi^\uparrow_{\uparrow+}&=\frac{k^1-ik^2}{x(1-x)}\phi(x,\vec{k}_\perp),\\
\psi^\uparrow_{\uparrow-}&=-\frac{k^1+ik^2}{1-x}\phi(x,\vec{k}_\perp),\\
\psi^\uparrow_{\downarrow+}&=\frac{1-x}{x}m\phi(x,\vec{k}_\perp),\\
\psi^{\uparrow}_{\downarrow-}&=0,
\end{align}
where
\begin{equation}
\phi(x,\vec{k}_\perp)=-\frac{\sqrt{2}e}{\sqrt{1-x}}\frac{x(1-x)}{\vec{k}_\perp^2+(1-x)^2m^2+x\lambda^2},
\end{equation}
with $\lambda$ being the photon mass parameter for infrared regularization. In our calculations, we take the massless limit $\lambda\to0$. The coefficients in front of $\phi(x,\vec{k}_\perp)$ are matrix elements of
\begin{equation}
\frac{\bar{u}_\sigma(k)}{\sqrt{k^+}}\gamma_\mu\epsilon_\lambda^{\mu *}\frac{u_{_{S_z}}(P)}{\sqrt{P^+}}.
\end{equation}
Similarly, the two-particle Fock state wave functions for the $S_z=-1/2$ electron are
\begin{align}
\psi^\downarrow_{\uparrow+}&=0,\\
\psi^\downarrow_{\uparrow-}&=\frac{1-x}{x}m\phi(x,\vec{k}_\perp),\\
\psi^\downarrow_{\downarrow+}&=\frac{k^1-ik^2}{1-x}\phi(x,\vec{k}_\perp),\\
\psi^\downarrow_{\downarrow-}&=-\frac{k^1+ik^2}{x(1-x)}\phi(x,\vec{k}_\perp).
\end{align}

As a common situation in perturbative theories beyond leading order, ultraviolet divergence happens in momentum integrals. We adopt the transverse dimensional regularization, and then the renormalization constant in Eq.~(\ref{fockexp}) is
\begin{equation}
Z=1+(\frac{3}{2}+2\mathrm{ln}\beta)(\frac{1}{\varepsilon}-1)\frac{e^2}{8\pi^2}-\frac{e^2}{16\pi^2},
\end{equation}
where $\beta$ is a cutoff parameter for light-front momentum fraction integration from $0$ to $1-\beta$ and does not appear in the final results, and $\varepsilon$ is the transverse dimensional regularization parameter defined as $d=2-2\varepsilon$, with $1/\varepsilon$ describing the logarithmic divergence in the ultraviolet region.

From the Fock state expansion~(\ref{fockexp}), one can directly get the canonical quantum numbers, fermion spin $S_f$, boson spin $S_b$, and total OAM $L$ for each Fock state. Their expected values are
{\allowdisplaybreaks
\begin{align}
S_f&=\frac{1}{2}-\frac{e^2}{16\pi^2},\label{sf}\\
S_b&=\frac{3e^2}{16\pi^2\varepsilon}-\frac{e^2}{8\pi^2},\label{sb}\\
L&=-\frac{3e^2}{16\pi^2\varepsilon}+\frac{3e^2}{16\pi^2}.
\end{align}
}The fermion and boson canonical (i.e., JM's) OAMs can be separately evaluated via the canonical OAM operator $\vec{r}_\perp\times\vec{k}_\perp$, where $\vec{r}_\perp$ is the transverse coordinate with respect to the transverse center of momentum as illustrated in~\cite{Lorce:2011kn}. Their expected values are
\begin{align}
L_f&=-\frac{e^2}{12\pi^2\varepsilon}+\frac{e^2}{12\pi^2},\label{lf}\\
L_b&=-\frac{5e^2}{48\pi^2\varepsilon}+\frac{5e^2}{48\pi^2},\label{lb}
\end{align}
and the sum of them is equal to the total OAM directly obtained from the quantum number $L$ of the Fock state. Obviously, the angular momentum sum rule is satisfied:
\begin{equation}
S_f+S_b+L_f+L_b=\frac{1}{2}.
\end{equation}

On the other hand, in Eqs.~(\ref{gffa})--(\ref{embplus}), we calculate the gravitational form factors at the zero momentum transfer limit. For the fermion
\begin{align}
A_f(0)&=1-\frac{e^2}{6\pi^2\varepsilon}+\frac{e^2}{8\pi^2},\\
B_f(0)&=\frac{e^2}{12\pi^2},
\end{align}
and for the boson
\begin{align}
A_b(0)&=\frac{e^2}{6\pi^2\varepsilon}-\frac{e^2}{8\pi^2},\\
B_b(0)&=-\frac{e^2}{12\pi^2}.
\end{align}
The $B_{f/b}(0)$ values have been obtained in~\cite{Brodsky:2000ii}. One can find that the $A$ form factors satisfy the momentum fraction sum rule:
\begin{equation}
A_f(0)+A_b(0)=1,
\end{equation}
and the $B$ form factors satisfy the anomalous gravitomagnetic moment sum rule:
\begin{equation}
B_f(0)+B_b(0)=0,
\end{equation}
which can be derived from the equivalence principle of gravity~\cite{Teryaev:1999su}. Therefore, they formally satisfy the relation
\begin{equation}
\frac{1}{2}[A(0)+B(0)]=S+L=\frac{1}{2},
\end{equation}
but, for each parton flavor, this relation is violated:
\begin{align}
\frac{1}{2}[A_f(0)+B_f(0)]&\neq S_f+L_f,\\
\frac{1}{2}[A_b(0)+B_b(0)]&\neq S_b+L_b.
\end{align}

This is not a surprise, because it is usually believed that the sum of gravitational form factors is related to the kinetic (i.e., Ji's) angular momentum, and is therefore different from the canonical one (i.e., JM's). Comparing the expressions, we find a finite difference between $[A_f(0)+B_f(0)]/2$ and $S_f+L_f$, i.e., $e^2/12\pi^2$. With explicit calculations, one can find that this difference has a dependence on the regularization procedures. In the cutoff regularization, the difference is $-e^2/96\pi^2$, and in the Pauli-Villars regularization, it is $-e^2/16\pi^2$ as given in~\cite{Burkardt:2008ua}. This regularization procedure dependence is not a serious problem in theoretical calculations and actually often happens, as demonstrated by the examples in~\cite{Brodsky:1997de}.

Since the kinetic and canonical decompositions share the same operator for the fermion spin term, this difference is usually attributed to the different part between the two fermion OAM operator definitions, i.e., the so-called potential angular momentum~\cite{Wakamatsu:2010cb}:
\begin{equation}
\Delta\hat{L}=\hat{L}^{\textrm{k}}_f-\hat{L}^{\textrm{c}}_f=-e\bar{\psi}\gamma^+\vec{r}_\perp\times\vec{A}_\perp\psi,
\end{equation}
where the superscripts k and c denote the kinetic and canonical definitions, respectively. But this has never been explicitly checked.

Here we perform an explicit calculation of the contribution from $\Delta\hat{L}$ by interpolating it between physical electron states. Unlike the operators we have evaluated above, it makes no contribution to the diagonal matrix elements in Fock space, but it has nonzero off-diagonal matrix elements between two Fock states that differ by a photon. After some algebra, we obtain its expected value as
\begin{equation}
\Delta L=-\frac{e^2}{8\pi^2\varepsilon}+\mathcal{O}(e^4).
\end{equation}
Obviously it does not compensate for the difference between the canonical angular momentum and half the sum of the two gravitational form factors. Apart from the $\Delta L$, a surface term, which may have nonvanishing contributions in plane waves, is neglected. Although constrained by the sum rules, the total surface contributions from the fermion and boson must vanish, but each component may have a nonzero expected value. Therefore, we need to calculate the expected value of the surface term. Taking the corresponding component of the surface generalized angular momentum tensor density operator for the fermion~\cite{Leader:2013jra}
\begin{equation}
\hat{M}^{\mu\nu\rho}_f=-\frac{1}{4}\partial_\lambda(x^{[\nu}\epsilon^{\rho]\mu\lambda\sigma}\bar{\psi}\gamma_\sigma\gamma_5\psi),
\end{equation}
one can obtain the operator of fermion longitudinal angular momentum from the surface term as
\begin{equation}
\hat{J}^{\textrm{surf}}_f=-\frac{1}{4}(\partial_\perp\cdot\vec{r}_\perp\bar{\psi}\gamma^+\gamma_5\psi+\vec{r}_\perp\cdot\partial^+\bar{\psi}\vec{\gamma}^\perp\gamma_5\psi).
\end{equation}
Sandwiching it between physical electron states, we get the expected value as
\begin{equation}
J^{\textrm{surf}}_f=-\frac{e^2}{12\pi^2\varepsilon}+\frac{5e^2}{48\pi^2}.
\end{equation}
We find that the angular momentum evaluated from the kinetic angular momentum operator---i.e., the sum of the canonical one, the $\Delta L$, and the surface term---does not equal half the sum of the two gravitational form factors for each constituent. Neither do they match in cutoff or Pauli-Villars regularizations. This result does not mean any decomposition version is invalid, but the relation to gravitational form factors is questionable, at least not order by order in perturbative theories. This is also the reason why the canonical and kinetic OAMs do not match with each other even in no gauge field models~\cite{Liu:2014npa,Lorce:2011kd} where the kinetic OAM is calculated from gravitational form factors or, equivalently, from the second moments of generalized parton distributions.

From the GIE procedure~\cite{Chen:2008ag,Lorce:2012rr},
both canonical and kinetic angular momentum decompositions are, in principle, measurable without breaking gauge invariance. The main challenge is to relate each term, especially the OAM, in the decomposition to physical observables. Some relations to a combination of helicity and transversity~\cite{Ma:1998ar}, the pretzelosity~\cite{She:2009jq}, and Sivers distributions~\cite{Bacchetta:2011gx} are proposed, but all of them are model dependent. Explicit perturbative QED calculations in this report indicate that the model-independent relation to gravitational form factors is not justified. At least it does not work order by order in perturbative theories. However, the measurement of these form factors is still of great interest, but with different physical significance. The DVCS process suggested in~\cite{Ji:1996ek} provides us an opportunity to get access to gravitational form factors through electromagnetic interactions. Recently, some relations to a generalized transverse momentum dependent parton distribution $F_{1,4}$ or the longi-unpolarized Wigner distribution $\rho_{_\textrm{LU}}$ are proposed~\cite{Lorce:2011kd,Lorce:2011ni}, though no feasible experimental approach has been found to extract this function. Therefore, we have no model-independent relations to measure OAM in an experiment at the moment, even though to extract OAM with model-dependent relations may provide us important information concerning the nucleon structure.

\acknowledgments{This work is supported by the National Natural Science Foundation of China (Grants No.~11035003, No.~11120101004 and No.~11475006).}


%

\end{document}